\documentclass[epj,final]{svjour}

\usepackage{latexsym}
\usepackage{url}
\usepackage{amsfonts}
\usepackage{amssymb}

\RequirePackage{graphicx}

\begin{document}
\title{Gravity lens critical test for gravity constants and dark sector}
\author{V.G. Gurzadyan\inst{1,2}, 
 A. Stepanian\inst{1}
}                     
%
%
\institute{Center for Cosmology and Astrophysics, Alikhanian National Laboratory and Yerevan State University, Yerevan, Armenia \and
SIA, Sapienza Universita di Roma, Rome, Italy}
\date{Received: date / Revised version: date}
%

\abstract{The recent study of the strong gravitational lens ESO 325-G004 \cite{Col} leads to a new possibility for testing General Relativity and its extensions. Such gravity lens observational studies can be instrumental for establishing a limitation on the precision of testing General Relativity in the weak-field regime and on the two gravity constants (the Newtonian and cosmological ones) as described in \cite{GS}. Namely, we predict a critical value for the involved weak-field parameter $\gamma_{cr}=0.998$ (for $M= 1.5\,\, 10^{11}\, M_{\odot}$ lens mass and $r=2\, kpc$ light impact distance), which remarkably does not depend on any hypothetical variable but is determined only by well measured quantities.  If the critical parameter $\gamma_{cr}$  will be established at future observations, this will mark the first discrepancy with General Relativity of conventional weak-field Newtonian limit, directly linked to the nature of dark sector of the Universe. 
} 

%
\maketitle

\section{Introduction}

High precision tests of the General Relativity (GR) and of its weak- and strong-field limits, along with of modified gravity models
have gained new content with establishing of the dominating dark sector of the Universe. The models proposed to describe the dark energy and dark matter observations are being constrained by means of those tests. The weak-field General Relativity is continuously being tested in ever increasing accuracy \cite{KEK,W} including the frame-dragging measurements with laser ranging satellites \cite{Ciu}.

The use of the strong gravity lens ESO 325-G004 \cite{Col} demonstrates the efficiency of the lens studies to constrain the weak-field GR in the intergalactic scales. Given the increasing accuracy and statistics of the gravity lensing observations (e.g. \cite{Cao}), those studies can become laboratories to test GR modifications along with the structure and dynamics of the dark matter dominated configurations.

The idea of our approach to this problem is the following. In recent studies \cite{GS,G17} we have shown that the Newton theorem on the identity of the gravity of a sphere and of a point mass located in its center provides a natural way for the weak-field modification of GR. Consequently, the constant $\Lambda$ appears both in the cosmological solutions describing the accelerated expansion and at galactic halo scales ({cf. \cite{G2}) as weak-field GR thus linking dark energy and galactic scales. 

Here we predict a critical value for the parameter of the parametrized post-Newtonian (PPN)
formalism \cite{W} $\gamma=0.998$ (normalized to given lens mass and light impact distance) which if observed at gravity lenses with proper significance will, for the first time, reveal the weak-field modification of GR differing from conventional Newtonian limit. It is important that the critical $\gamma$ does not depend on any open parameter but involves only the fundamental constants and well measured quantities. 

\section{Newton's theorem and gravity lensing with $\Lambda$}

In \cite{GS} it is shown that from the Newton's theorem the weak-field limit of GR follows involving the cosmological constant $\Lambda$, so that the metric tensor components have the form
\begin {equation} \label {mod}
g_{00} = 1 - \frac{2 G m}{r c^2} - \frac{\Lambda r^2}{3};\,\,\, g_{rr} = (1 - \frac{2 G m}{r c^2} - \frac {\Lambda r^2}{3})^{-1}.
\end {equation} 
While this metric (Schwarzschild -- de Sitter) was obviously known before (e.g. \cite{R,N}), with the $\Lambda$ introduced by Einstein to get a static cosmological model, the motivation based on the Newton's theorem essentially differs from that and ensures the Newton's potential with $\Lambda$ term as the weak-field limit of GR.   

Namely, as follows from the consideration of the general function for the force satisfying Newton's theorem on the identity of sphere's gravity and that of a point of the sphere's mass situated in its center. That function besides the $r^{-2}$ term contains also a second term \cite{G85} 
\begin{equation}
f(r)= Ar^{-2} + \Lambda r.
\end{equation}   
When the modified Newtonian law (for the potential) is taken as weak-field limit of GR, one has the constant $\Lambda$ as a second gravity constant along with the classical Newtonian constant $G$ \cite{GS}. Thus, the second constant acts as cosmological constant in the solutions of Einstein equations, at the same time enters the low-energy limit of GR.

Turning to the strong lensing and following \cite{Col}, it is convenient to deal with the parameter representing the ratio 
\begin{equation}
\gamma=\Psi/\Phi
\end{equation}
of the two perturbing functions - of the Newtonian potential $\Phi$ and curvature potential $\Psi$ - entering the weak field metric
\begin{equation}
ds^2= (1+ 2\Phi) c^2 dt^2 - (1-2\Psi) dr^2 - r^2 d\Omega^2.
\end{equation}
For GR $\gamma =1$, obviously. 

In the weak-field limit of GR following from the Newton's theorem Eq.(\ref{mod})  \cite{G17,GS}, we obtain
\begin{equation}
\Phi=\Psi= -\frac{GM}{r c^2} - \frac{\Lambda r^2}{6}.
\end{equation}

To find out $\gamma$  the authors of \cite{Col} introduced the lense dynamical mass
\begin{equation}
M_{dyn}= \frac{1+\gamma}{2} M, 
\end{equation}
where consequently the light bending angle in the vicinity of a mass of surface density $\Sigma$ is achieved
is:
\begin{equation}
\alpha= \frac{2G(1+\gamma)}{c^2}\int d^2 x' \Sigma(x')\frac{x-x'}{|x-x'|^2}.
\end{equation}
Upon the analysis of the observational data of ESO 325-G004 they finally obtain the value $\gamma \simeq 0.97 \pm 0.09$.

Now, within our approach of the weak field metric Eq.(\ref{mod}) the bending angle
will be \cite{Is}
\begin{equation}
\alpha= \frac{4 G M}{c^2 r} - \frac{\Lambda c^2 r^3}{6 G M}.
\end{equation}
Here, important difference arises between Newtonian and $\Lambda$-modified case. Namely, the authors in \cite{Col} have obtained 
\begin{equation}
\alpha = 2 (1+\gamma) \frac{GM}{c^2 r}. 
\end{equation}
Comparing this with Eq.(8),  we get for the $\gamma$-parameter}
\begin{equation}
\gamma= 1-\frac{\Lambda c^4 r^4}{12G^2 M^2}.
\end{equation}
Inserting the current value of the cosmological constant e.g. that of the Planck satellite \cite{Pl} $\Lambda=1.11 \times 10^{-52} m^{-2}$ we obtain 
\begin{equation}
\gamma_{cr} = 1- 0.002 =0.998\,\,(\frac{M}{1.5 \,\,  10^{11}\, M_{\odot}})^{-2}(\frac{r}{2\,kpc})^4,
\end{equation}
where the data of \cite{Col,Sm} for ESO 325-G004, i.e. the Einstein radius and the estimated mass inside that radius, were used for the normalization. Obviously, the normalization and hence the precise numerical value of $\gamma_{cr}$ will vary from one lens to another, and the principal point is the existence of a well defined parameter enabling to reveal the weak-field limit of GR with the lensing effect. 

The other key point of using  ESO 325-G004 is that, within other available observational means it is technically impossible to detect the contribution of $\Lambda$ term in the gravitational lensing. For example, for the same effect within the Solar System the value of $\gamma_{cr}$ will be approximately (1- 9.6 $\times$ 10${}^{-25})$. Note, that a limitation on $\gamma$ also will emerge due to the proper motion of the lens for it affects the measured value of $\gamma$ as shown in \cite{Kop}.

\section{Conclusions}

The accurate measurements of strong lensing of extragalactic objects provide important means for the study of profound cosmological problems. The recent study of the lensed object ESO 325-G004 \cite{Col} (cf. previous studies \cite{Cao,Bol,Sch}) enables a remarkable testing of General Relativity in extragalactic scales. The observational surveys of lensing will definitely proceed further with ever increasing precision and statistics which will enable to improve the available accuracy of the value of the weak-field parameter $\gamma$.

In view of that, here we derive a critical $\gamma_{cr}$ for the strong lensing of extragalactic objects which can be informative for gravity theories. It is remarkable that, $\gamma_{cr}$ does not depend on any hypothetical parameter of modified gravity models (coupling constant, scalar field mass, etc) but is determined entirely by measured physical quantities. Since the needed accuracy for measuring of $\gamma_{cr}$ seems not principally unreachable given the variability range of the lense mass and light impact scale, due to $\gamma_{cr}$ the gravity lens measurements will get similar importance as the renown Solar eclipse of 1919 which enabled to distinguish GR from the classical Newtonian gravity. Thus, for the first time the detection of a discrepancy with the conventional General Relativity can become feasible, with further intriguing relation to dark sector.  While the breakthrough study \cite{Col} reveals the possibility of obtaining of $\gamma$, the observations of more distant objects and hence the needed accuracy for $\gamma$ certainly    
is a matter of future advances; however, let us recall the classical example, i.e. Einstein's scepticism as regards observing gravitational lenses \cite{E}. 

\section{Acknowledgement}
We are thankful to the referee for helpful comments and S. Mirzoyan for discussions. AS acknowledges the Abdus Salam International Centre for Theoretical Physics Affiliated Center program AF-04 for financial support.


\begin{thebibliography}{00}


\bibitem{Col} T.E. Collett, et al, Science, 360, 1342 (2018)
\bibitem{GS} V.G. Gurzadyan, A. Stepanian, Eur Phys. J. C, 78, 632 (2018)
\bibitem{KEK} S. Kopeikin, M. Efroimsky, G. Kaplan, Relativistic celestial mechanics of the solar system, (Wiley, 2001)
\bibitem{W} C.M. Will, Living Rev. Relativ., 17, 4 (2014)
\bibitem{Ciu} I. Ciufolini, et al, Eur. Phys. J. C 76, 120 (2016)
\bibitem{Cao} S. Cao,  et  al., ApJ , 835, 92 (2017)  
\bibitem{G17} V.G. Gurzadyan,  arXiv:1712.10014 (2017)
\bibitem{G2} V.G. Gurzadyan, et al, A \& A, 609, A131 (2018) 
\bibitem{G85} V.G. Gurzadyan, Observatory,  105, 42 (1985)
\bibitem{R} W. Rindler, Relativity, Special, General, and Cosmological (Oxford University Press, 2006)
\bibitem{N} M. Nowakowski,  Int.J.Mod.Phys. D10, 649 (2001)  
\bibitem{Is} M. Ishak, W. Rindler, J. Dossett, J. Moldenhauer, C. Allison, MNRAS, 388, 1279 (2008) 
\bibitem{Pl} P.A.R. Ade, et al,  A\&A 594, A13 (2016)
\bibitem{Sm} R.J. Smith,  J.R. Lucey, MNRAS, 434, 1964 (2013)
\bibitem{Kop} S. Kopeikin, MNRAS, 399, 1539  (2009)
\bibitem{Bol} A.S. Bolton, S.A. Rappaport, S. Burles, Phys. Rev. D, 74, 061501 (2006)
\bibitem{Sch} J. Schwab, A.S. Bolton, S.A. Rappaport, ApJ, 708, 750 (2010)
\bibitem{E} A. Einstein, Science, 84, 506 (1936)  




\end{thebibliography}
\end{document}